\documentstyle[12pt]{article}
\font\titolo=cmbx12 scaled\magstep2
\font\titolino=cmbx12
\font\tsnorm=cmr12

\font\tscorsp=cmti10

\def\NPB{{\em Nucl. Phys. }}
\def\PLB{{\em Phys. Lett. }B }  

\def\PRD{{\em Phys. Rev. }D } 
 
\def\MPLA{{\em Mod. Phys. Lett. }A} 
 
\def\z{Z\kern -4.6pt Z}
\def\dx{\int d^2x\ \sqrt{-g}\ }
\def\xp{x^+}\def\xm{x^-}
\def\l{\lambda}
\def\ha{{1\over 2}}

\def\ha{1\over 2}
\def \ma{{2 M\over \l}}
\def\na{\nabla}

\def\f{\phi}

\def\m{\mu}
\def\n{\nu}
\def\h{\hat}
\def\gmn{g_{\m\n}}
\def\hgmn{\hat \gmn}
\def\dx{\int d^2x\ \sqrt{-g}\ }
\def\dhx{\int d^2x\ \sqrt{-\hat g}\ }      
\def\r{\rho}
\def\s{\sigma}

\def\t{\tau}

\def\be{\begin{equation}}
\def\ee{\end{equation}}
\def\bea{\begin{eqnarray}}
\def\eea{\end{eqnarray}}
\def\bc{\begin{displaymath}}
\def\ec{\end{displaymath}}
\def\lb{\label}
\begin{document}
\pagestyle{empty}
\null
\vskip 5truemm
\begin{flushright}
INFNCA-TH9623\\
October 1996
\end{flushright}
\vskip 15truemm
\begin{center}
{\titolo CONFORMAL EQUIVALENCE OF 2D DILATON}
\end{center}
\begin{center}
\titolo{GRAVITY MODELS}
\end{center}
\vskip 15truemm
\begin{center}
{\tsnorm Mariano Cadoni}
\end{center}
\begin{center}
{\tscorsp Dipartimento di Scienze Fisiche,  
Universit\`a  di Cagliari,}
\end{center}
\begin{center}
{\tscorsp Via Ospedale 72, I-09100 Cagliari, Italy.}
\end{center}
\begin{center}
{\tscorsp and}
\end{center}
\begin{center}
{\tscorsp  INFN, Sezione di Cagliari.}
\end{center}
\vskip 19truemm
\begin{abstract}
\noindent
We investigate the behavior of generic, matter-coupled, 2D dilaton 
gravity theories under dilaton-dependent Weyl rescalings of the metric.
We show that physical observables associated with
2D black holes, such as 
the mass, the temperature and the flux of Hawking radiation are 
invariant  under the action of  both  Weyl transformations and 
dilaton reparametrizations. The field theoretical and geometrical 
meaning of these invariances is discussed.
\end{abstract}
\vfill
\hrule
\begin{flushleft}
{E-Mail: CADONI@CA.INFN.IT\hfill}
\end{flushleft}
\vfill
\eject
\pagenumbering{arabic}
\pagestyle{plain}

The recent flurry of activity on two-dimensional (2D) black hole physics 
\cite{ST1} 
even though it
has not succeeded in finding a definite answer to   challenging 
questions such as the ultimate fate of black holes or the loss of 
quantum coherence, has enabled us to gain considerable knowledge
on the subject.
Among other things, we have got a strong indication that a 
consistent description of black holes at the semiclassical or 
even quantum level requires us to treat the matter and the 
gravitational degrees of freedom on the same footing.    
Considerable progress has been achieved by considering 2D dilaton 
gravity models from this purely field theoretical point of view, for 
example as a non-linear $\s$-model \cite {KST, RST}, as a 2D conformal 
field theory \cite {BC&A}  or in 
the gauge theoretical formulation \cite {JA}. 

One serious problem of this kind of approach is the difficulty in 
giving a geometrical interpretation to some field theoretical concepts.
For example, from a purely field theoretical point of view, performing 
dilaton-dependent Weyl rescalings of the metric in the 
2D dilaton gravity action
should give us 
equivalent models, being these transformations nothing but 
reparametrizations of the field space. The space-time interpretation of 
this equivalence presents, however, some problems. Though the causal 
structure of the 2D space-time does not change under Weyl 
transformations, 
geometrical objects such us the scalar curvature of the space-time or 
the equation for the geodesics do change. This discrepancy has 
generated a lot a confusion on the subject. Some authors have assumed 
explicitly or 
implicitly this equivalence to hold and used it to simplify the  
description of the general model \cite {MK,GKM, BL} or even to argue about 
the existence  of 
Hawking radiation in the context of the 
Callan-Giddings-Harvey-Strominger model (CGHS) \cite {JA,CGHS}. 
Other authors, 
focusing on the space-time 
interpretation of the gravitational degrees of freedom, 
have pointed out the non-equivalence of 2D dilaton 
gravity models connected by 
Weyl rescalings of the metric 
\cite {CGHS, LVA}.

In this paper we analyze in detail the role of conformal 
transformations
of the 
metric in the context of generic, matter-coupled, 2D dilaton gravity 
theories. We prove a conjecture reported in a previous paper \cite 
{CA}, and 
based on  previous results for the CGHS model \cite {CM3},
namely that  physical observables for 2D black holes, 
such us 
the mass, the temperature and the flux of Hawking radiation, are invariant under 
dilaton-dependent Weyl rescalings of the metric. Moreover, 
we  show 
that the same observables are also invariant under reparametrization 
of the dilaton  field.
 
The most general action of 2D dilaton gravity conformally coupled to 
a set on $N$ matter scalar fields has the form \cite {BL, MA}    
\be
S[g,\f,f] ={1\over 2\pi}\dx\left[D(\f) R[g]+H(\f)(\nabla \f)^{2}+
\l^2 V(\phi)-{\ha} \sum^N_{i=1}(\na f_i)^2 \right],
\lb{e1}
\ee
where $D,H,V$ are arbitrary functions of the dilaton $\f$ and $\l$ is 
a constant. 
Let us consider the following Weyl transformations of the metric
\be
\gmn= e^{P(\f)} \hgmn,
\lb{e2}
\ee
for the moment we constrain the  form of the function $P$ only by 
requiring  the transformation (\ref{e2}) to be non-singular and invertible
in the  range of variation of the dilaton.
Arguments related to the geometrical interpretation of the 
transformation (\ref{e2}) 
will impose some  additional restrictions on the  
form of the function $P$. We will come back to this point 
later on this paper.
Whereas the matter part of the action (\ref{e1}) is invariant under 
the transformation (\ref{e2}), the gravitational part is not. 
Nevertheless, the latter  maintains its form under these transformations, in 
fact, modulo a total derivative we have
\begin{displaymath}
 S[g,\f] \to S[{\h g},\f]={1\over 2\pi}\dhx\left[{\h D}(\f) R[{\h g}]+{\h H}(\f)
({\h \nabla} \f)^{2}+\l^2 {\h V}(\phi) \right],
\end{displaymath}
where the new functions ${\h D},{\h H}, {\h V}$ are related to the old ones 
through the  transformation laws ($'=d/d\f$)
\be\lb{e4}
{\h D}=D,\qquad
{\h H}=H +D'P',\qquad
{\h V}=e^{P}V.
\ee
Under dilaton reparametrizations $\f=\f(\tilde \f)$,   $V$ and $D$ behave 
as scalars, whereas $H$ transforms as 
\be
\tilde H(\tilde\f)=H(\f) \left( {d\f\over d \tilde\f}\right)^{2}.
\lb{e5}
\ee
The transformation laws (\ref {e2}), (\ref{e4}) and (\ref{e5}) 
enable us to find out how the physical parameters characterizing the 
solutions of the theory transform under  Weyl transformations (\ref {e2})
and dilaton reparametrizations. 

Let us begin with the mass of the solutions.
R. B. Mann has shown that for the generic  theory defined by the 
action (\ref{e1}), one can define the conserved quantity \cite {MA} 
\be
M={F_{0}\over 2} \left[\int^{\f}dD  V\exp\left(-\int d\t{H(\t)\over 
D'(\t)}\right) -(\na D)^{2} \exp\left(-\int d\t{H(\t)\over 
D'(\t)}\right)\right],
\lb{e6}
\ee
where $F_{0}$ is a constant. $M$ is constant whenever the equation of 
motion are satisfied and,  in this case, it can be interpreted as the mass 
of the solution. Using  Eqs. (\ref{e2}), (\ref{e4}), 
(\ref{e5}), one can easily demonstrate that the mass $M$ given by
the expression (\ref{e6}) is invariant under both Weyl transformation 
 and dilaton reparametrizations.
Notice that this invariance means not only that two conformally  related 
solutions have the same mass, but also 
that the off-shell quantity (\ref{e6}) is Weyl-invariant. 

The Hawking temperature associated with a generic black hole solution 
can be defined as the inverse of  the periodicity of the Euclidean time 
necessary to remove the conical singularity at the event horizon. 
To perform this calculation we need an explicit form for the black 
hole solutions of 2D dilaton gravity. The generic static 
solutions in the conformal frame in which ${\h H}=0$,
have already been found in Ref. \cite {MK, GKM}, 
\be
{\hat{ds}}^{2}= a^{2}\left(\h J-{2M\over \l^{2}F_{0}}\right)dt^{2}+
a^{-2}\left(\h J-{2M\over \l^{2}F_{0}}\right)^{-1}dr^{2}, 
\quad D(\f)={\l \over a} \, r,
\lb{e8}
\ee
where $d\h J/d{\h D}={\h V}$, $a$ is an arbitrary integration constant and $M$ 
is the mass of the solution given by Eq. (\ref{e6}).
The static solutions in the generic conformal frame can be easily 
obtained from these solutions using  Eq. (\ref {e2}) with 
$P=-\int^{\f}d\tau \left[H(\tau)/ D'(\tau)\right]$,
\be
ds^{2}=\exp\left(-\int^{\f} d\tau {H(\tau) \over D'(\tau)}\right)
{\hat {ds}}^2.
\lb{e9}
\ee
A straightforward calculation gives  for the Hawking temperature 
associated with an  event horizon of the solution (\ref{e9}),  located at 
 $\f=\f_{0}$, 
\be
T={\l a\over 4\pi} V(\f_{0})\exp\left(-\int^{\f_{0}} d\tau {H(\tau) 
\over D'(\tau)}\right).
\lb{f1}
\ee
The temperature is invariant  both under Weyl transformations
 and dilaton reparametrizations. This can be easily checked 
 using Eqs. (\ref{e4}) and (\ref{e5}) in Eq. (\ref{f1}) and taking 
into account that the transformations (\ref{e2}) do not change 
the position of the event horizon, because by assumptions they are
everywhere non-singular and invertible.

In Eq. (\ref{e8}) and in the expressions (\ref{e6}), 
(\ref {f1}) for the mass and the temperature appear two arbitrary 
constants $a$ and $F_{0}$. As already noted in Ref. \cite {LVA}  for a 
particular 
class of 2D dilaton gravity models, their presence is related to the 
arbitrariness in defining the asymptotical behavior of the metric or, 
from a physical point of view, to the way an asymptotical observer 
measures lengths and masses.
Our prove does not rely on the way one fixes this arbitrariness, 
nevertheless, 
it is useful to have a definite and general prescription to fix it.
The proposal of Ref. \cite {LVA}  seems to us too much 
model-dependent; 
we will use 
here a different prescription.
First of all, we need a notion of asymptotic region (spatial 
infinity) for our space-time, which is Weyl-invariant. As already noted 
in a previous 
paper \cite {CA}, the dilaton $\f$ gives a coordinate-independent notion of 
location and can therefore be used to define the asymptotic region, 
the singularities and the event horizon of our 2D space-time. 
Moreover, the natural coupling constant of the theory is $D^{-1/2}$
so that we have a natural division of our space-time in a 
strong-coupling region ($D=0$) and a weak-coupling region ($D=\infty$).
These considerations limit the range of variation of the function $D$ 
to $ 0\le D<\infty$ and enable us to identify the weak-coupling region 
$D=\infty$ with the asymptotic region of our space-time \cite {CA}.
Moreover, this notion of 
location is Weyl-invariant because the term 
$\sqrt{-g}D R[g]$ in the action (\ref{e1}) is transformed by 
(\ref{e2}) into $ \sqrt{-\h g}D R[\h g]$.  

In the conformal gauge 
\be\lb{f12}
ds^{2}=-e^{2\rho}dx^{+}dx^{-},
\ee
 using a Weyl transformation 
(\ref{e2}) one can always  put the solution (\ref {e9}) into the  
form

\be 
e^{2\rho}= a^{2}\left(1- {2M\over \l F_{0}K}\right),\qquad 
K=\int^{\f} dD V \exp\left(-\int d\tau{H(\tau)\over D'(\tau)}\right).
\lb{f2}
\ee
This form of the solution can be used to fix the values of the parameters 
$a$ and $F_0$.
Using arguments similar to those of Ref. \cite {CA}, one can show that a 
black hole interpretation of the solution (\ref{f2}) requires $K\to 
\infty$ for $D\to \infty$. The condition that the metric (\ref{f2}) has
asymptotically a
Minkowskian form fixes now $a=1$.
The general solution admits a Killing vector of the form \cite {MA} 
\begin{displaymath}
\zeta ^{\m}=\epsilon^{\m\n}\na _{\n}{\cal F},\qquad 
{\cal F}= F_{0}\int^{\f} dD \exp\left(-\int d\tau {H(\tau)\over 
D'(\tau)}\right).
\end{displaymath}
The constant $F_{0}$ can be  fixed to $F_{0}=1/\l$ by requiring 
that  in the 
conformal frame in which the metric has the form (\ref{f2}), the 
norm of the Killing vector approaches, for
$D\to \infty$,  the value $-1$.

To discuss the Hawking effect we need to be sure that the solutions 
we are dealing with really represent black holes. 
Our previous discussion does not rely heavily on the notion of black 
hole. The mass formula (\ref{e6}) holds for every solution of the 
theory, whereas the temperature  (\ref{f1}) is a local-defined quantity, 
which does not care if the space-time has the global features of a 
black hole. In view of the discussion of  Ref. \cite {CA}, 
one  expects that the  form of the 
functions $D, H,V$ has to be constrained in order to be sure that 
the solution (\ref{e9}) is a black hole. However, the discussion of Ref.
\cite {CA}  cannot be extended trivially to the present context. 
In  Ref. \cite {CA}  we used  the scalar 
curvature $R$ 
to define the singularities and the asymptotic behavior of the 
space-time. $R$ is not  Weyl rescaling 
invariant  and cannot be taken as a good quantity for a conformal
invariant characterization of black holes. 
Here, we will not tackle the problem in this general setting, but we 
will consider the black hole solutions in the particular 
conformal frame in which the metric is asymptotically 
Minkowskian and has, therefore, the form (\ref{f2}). In this conformal frame, 
the ground state solution  $M=0$  coincides with Minkowsky space.
Assuming that the black holes exist in any 
conformally related frame, we will show that the result for the 
Hawking radiation  is invariant under conformal transformations 
of the metric.

In the conformal frame defined by Eq. (\ref{f2}), the scalar 
curvature  of 
the black hole space-time is
\begin{displaymath}
R= 2M \l K{d^{2}\over dD^{2}}\ln K.
\end{displaymath}
We require that the $M\neq 0$ solutions behave asymptotically as the 
ground state solution, i.e.,  $R\to 0$ for $D\to \infty$.
This singles out three main classes of 2D dilaton gravity models,
according to the asymptotical, $D\to \infty$, behavior of 
the function $K$:
\bea \lb{f5}
K&\sim &D^{\alpha}, \qquad 0<\alpha<2,\nonumber\\
K&\sim& \gamma \ln D, \qquad 0<\gamma<\infty,\\
K&\sim&e^{\beta D}, \qquad 0<\beta<\infty\nonumber.
\eea
The first  class of models has already been   found  and  discussed 
in Ref. \cite {CA}.  
Our discussion, including the Hawking effect, holds also for models 
with $\alpha=2$.  In this case 
the solutions describe space-times that are asymptotically anti-de Sitter. 

There are various ways to analyze the Hawking effect.  Here, we will 
 use the relationship between Hawking radiation 
and quantum anomalies \cite {CS, CGHS, AS,CA}. It is well-known that in 
quantizing the scalar matter fields $f$ in a fixed background geometry 
 the Weyl rescaling  or/and part of the diffeomorphism  invariance of the 
classical action for the matter fields has to be 
explicitly broken. The quantization procedure has two sources of ambiguity. 
First, one can 
decide to preserve at the semiclassical level either the 
diffeomorphism or the Weyl rescaling invariance \cite {KMM, JA1, ABS} 
( for sake of 
simplicity we do not consider 
here the case in which both symmetries are broken). Second, if one 
decides to preserve diffeomorphism invariance, one has still the freedom 
of 
adding local, covariant, dilaton-dependent counterterms to the 
semiclassical action \cite {ST, BC&A, RST}. The nature of these ambiguities is 
particularly 
clear in the path integral formulation. By choosing the 
diffeomorphism-invariant measure \cite {DK} 
\be \lb{f6}
\int{\cal D}f_{i} \exp\left(i\dx f_{i}^{2}\right)=1,
\ee
one breaks explicitly the Weyl invariance of the classical matter action,
and introduces an ambiguity related 
to the choice of the metric to be used in the measure. One is allowed 
to use in Eq. (\ref{f6}) the metric $\gmn$ or a Weyl-rescaled metric 
$\hgmn$. The corresponding semiclassical actions differ 
one from the 
other for the presence of local, covariant, dilaton-dependent counterterms.
On the other hand, by choosing the Weyl-invariant 
measure \cite {JA1, ABS} 
\be \lb{f7} 
\int{\cal D}f_{i} \exp\left(i\int d^{2 }x f_{i}^{2}\right)=1,
\ee
one breaks part of the diffeomorphism invariance  of the 
classical action, but  
there is no ambiguity associated with the choice of the metric to be used in the 
measure (the measure (\ref{f7}) does not depend on the metric).
It has already been shown for a particular 2D dilaton gravity model 
(the CGHS model) that, though the form of the Weyl anomaly depends on the 
choice of the measure, the flux of the Hawking radiation 
does not \cite {AS}.
Here, we will not only show that this is true for a generic 2D dilaton 
gravity model but also that the result for the Hawking 
radiation is independent of the metric used in the measure (\ref{f6}), 
which is equivalent to prove the invariance of the Hawking 
radiation rate under 
the Weyl rescaling (\ref {e2}). To be more precise, in Ref. \cite {AS}  the 
measure and the trace 
anomaly is parametrized by a real parameter $k$. The two cases we discuss 
here correspond respectively to $k=1$ and $k=0$. We expect that our 
considerations can be trivially extended to arbitrary values of $k$.

Let us first consider a  measure defined by Eq. (\ref {f6}). 
The semiclassical effective action is diffeomorphism-invariant and
in the conformal 
frame where the metric is asymptotically Minkowskian,  it is 
 given by
\be \lb{f8}
S_{sc}=S_{cl}-S_{lp},
\ee
where $S_{cl}$ is the classical action (\ref{e1}) and 
 $S_{lp}$ is the usual non-local Liouville-Polyakov action
\begin{displaymath}
S_{lp}={N\over 96 \pi} \int d^{2 }x \sqrt{-\bar{g}}R[\bar g]
\bar \na^{-2} R[\bar g],
\end{displaymath}
where  the notation $\bar g$ has been used  in order to avoid 
confusion with the metric in the generic conformal 
frame.

The semiclassical action has its "minimal" 
Liouville-Polyakov form, with no dilaton-dependent counterterms 
present, exactly in the conformal frame where the solutions are 
asymptotically Minkowskian. This fact follows from very simple physical 
requirements.
Dilaton-dependent counterterms are forbidden if one requires that
the expectation value of the stress-energy tensor vanishes when 
evaluated for Minkowsky   
space (the $M=0$ ground state solution of our models).
Under a Weyl transformation (\ref{e2}) the Liouville-Polyakov action acquires 
local, dilaton-dependent terms that have the same form as those 
already present in the action (\ref{e1}). These terms depend on the 
form of the function $P(\f)$ in Eq.  (\ref{e2}), so that -as expected- 
the trace anomaly depends on the particular conformal frame chosen.
Using the equation ${\bar g}_{\m \n}= \gmn\exp\left(\int d\tau [{H(\tau)/ 
D'(\tau)}] -\ln K\right) $ in the expression (\ref{f8}), one easily finds the 
form of the semiclassical action in the generic conformal frame:
\bea\lb{g1}
S_{sc}[g]&=&S_{cl}[g] - S_{lp}[g]\nonumber\\ 
 &-& {N\over 96 \pi}\dx\left[ 2\left(\ln K - 
\int^{\f} d\tau{H(\tau)\over 
D'(\tau)}\right)R[g]-\left({K'\over K}-{H\over D'}\right)^{2 }(\na 
\f)^{2}\right].
\eea
The black hole radiation can now  be studied along the lines of 
Ref. \cite {CA}, working in the conformal gauge (\ref{f12}) and 
considering a black hole 
formed by collapse of a $f$-shock-wave, traveling in the $x^{+}$ 
direction and described by a classical stress-energy tensor $T_{++}=M\delta(x^{+}-x^{+}_{0)}$.
The classical solution describing the collapse of the 
shock-wave, for $\xp\le\xp _0$, is given by
\be\lb{g3}
e^{2\r}=K\exp\left(-\int^{\f} d\tau {H(\tau)\over 
D'(\tau)}\right), \qquad
\int^\f{d\t\over K(\t)}={\l\over 2}(\xp-\xm),
\ee
and, for $\xp\ge\xp_0$, it is given by 
\begin{displaymath}
e^{2\r}=\exp\left(-\int^{\f} d\tau {H(\tau)\over 
D'(\tau)}\right) \biggl(K-\ma\biggr)F'(\xm),
\end{displaymath}
\be\lb{g4}
\int^\f{d\t\over K(\t)-\ma}={\l\over 2}\left[\xp-\xp_0 -F(\xm)\right],
\ee
\be\lb{g5}
F'(\xm)={dF\over d\xm}=\biggl({K\over 
{K-\ma}}\biggr)_{\xp=\xp_0}.
\ee
The next step in our semiclassical  calculation is to use the 
effective action (\ref{g1}) to derive the expression for the quantum 
contributions of the matter to the stress-energy tensor. The flux of 
Hawking radiation across spatial infinity is given by $<T_{--}>$ 
evaluated on 
the asymptotical $D=\infty$ region.
For the class of models in Eq. (\ref{f5}) a 
straightforward calculation, which follows closely that of 
Ref. \cite {CA}, 
leads to
\be
\lb {g51}
 < T_{--}>_{as}={N\over 24}{1\over (F')^2}\{F,\xm\},
\ee
where $\{F,\xm\}$ denotes the Schwarzian derivative of the function 
$F(\xm)$.
This is a Weyl rescaling and dilaton reparametrization invariant 
result for the Hawking flux. 
In fact the function $F(\xm)$ is defined entirely in terms of the 
function $K(\f)$ (see Eq. (\ref{g5})), which in turn is invariant 
under both 
transformations (see Eq. (\ref{f2}).
Though the trace anomaly is Weyl rescaling 
dependent, the Hawking radiation seen by an asymptotic observer is 
independent of the particular conformal frame chosen.      
When the horizon $\f_{0}$ is approached, the Hawking flux reaches the thermal 
value 
\be\lb{g52}
< T_{--}>_{as}^{h}={N\over 12 }{\l^{2}\over 
16}\left[ V(\f_{0})\exp\left(-\int^{\f_{0}} d\tau {H(\tau)\over 
D'(\tau)}\right)\right]^{2},
\ee
which is the result already found in Ref. \cite {CA}, written 
in a manifest Weyl rescaling and dilaton 
reparametrization invariant form.

Let us now consider a Weyl-invariant measure defined by Eq. (\ref {f7}). 
In this case there is no Weyl anomaly, no ambiguity associated with 
the choice of the conformal frame and therefore there is no need to 
introduce dilaton-dependent counterterms in the effective action. 
The Hawking effect is a 
consequence of the explicit breaking of  diffeomorphism 
invariance. $T_{\m\n}$ does not transform as a tensor
 under general coordinate 
transformations $y^{\m}=y^{\m}(x^{\m})$, but it transforms  
as follows \cite {AS}
\begin{displaymath}
T_{\m\n}^{(y)}=(T_{\alpha\beta}^{(x)}+\Delta_{{\alpha\beta}}){dx^{\alpha}\over 
dy^{\m}}{dx^{\beta}\over 
dy^{\n}}.
\end{displaymath}
Applied to the shock-wave solution (\ref{g3}),(\ref{g4}), this transformation law 
gives the flux of Hawking radiation entirely in terms of the Schwarzian 
derivative of the conformal map $x^{-}\to F(\xm)$, with $F$ given by 
Eq. (\ref{g5}), i.e. it reproduces our previous result (\ref{g51}).

Let us now come to a central question: What is the 
physical meaning of the equivalence we have found? 
Does it imply the complete physical 
equivalence (at least at the semiclassical level) of
2D dilaton gravity models connected by Weyl transformations?
The answer to these questions is rather complex. A good starting point 
for trying to find an answer is to ask ourself the opposite question:
 What are the 
features of the models that {\sl do change}  under a Weyl rescaling of 
the metric?
We have already noted that because the scalar curvature of the 
space-time changes under Weyl transformations, in general 
the structure of the space-time singularities is not preserved by 
such transformations. Also the notion of geodesic motion depends on 
the choice of the conformal 
frame. In Fact, the geodesic equation
is not invariant under the transformation (\ref{e2}), but acquires 
terms depending on the derivatives of the function $P$. As a 
consequence a space-time that is geodesically complete in the range
$0\le D(\f)<\infty$ can be mapped by the transformation (\ref{e2}) 
into a space-time which is not geodesically complete in the same 
range of variation for the dilaton.

The model discussed in Ref. \cite {CM3}  gives a nice example of what happens.
In Ref. \cite {CM3}  has been shown that the black hole solutions of 
the CGHS model can be mapped by a 
conformal transformation (\ref{e2}) into the black hole solutions of a 
conformally  related model, leaving invariant the  
 values of mass, temperature and Hawking flux.
The Weyl rescaled solutions (which describe essentially Rindler 
space-time) differ from the CGHS ones in two respects. First, whereas 
the black hole solutions of the CGHS model 
have a curvature singularity, 
the Weyl-rescaled ones describe a space-time with no curvature singularities. 
The structure of the singularities  of the CGHS model is however preserved by considering
the dilaton on the same footing as the metric, imposing a boundary on 
the space-time in the strong-coupling region of the theory. Second, 
whereas the ground state of the CGHS model (the so called linear 
dilaton vacuum) is a geodesically complete space-time, the ground 
state of the conformally related model is not  geodesically complete
in the allowed range of variation for the dilaton.

The answer to the previous questions depends on the features of 
the model we are interested in. After all, 2D dilaton gravity models 
are just toy models for studying  4D gravity in a simplified 
context. Differently from the 4D case,  where  geometrical objects such 
as the metric or the curvature have a direct physical meaning,  
in the 2D case these objects acquire  a physical significance only  
through their 
relation with the 4D problem. There are examples in 
which 2D space-times with asymptotical anti-de Sitter behavior can be 
used to model asymptotically flat 4D black holes near extremality 
\cite {CM1}.
If geometrical features of the 2D model such  as the curvature or the 
geodesic completeness of the space-time are crucial for our problem, we 
will consider models related by Weyl transformations as non-equivalent.
On the other hand, if these geometrical features are irrelevant 
because we want to treat the gravitational degrees of freedom on the same 
footing as the matter degrees of freedom or because  
our problem is focused on physical observables such as masses or 
temperatures, we 
can  regard the former models as equivalent.

In this discussion the form of the function $P(\f)$ in Eq. (\ref{e2}) 
plays a crucial 
role. Until now we have assumed, generically, that $P$ is such that the 
transformation (\ref{e2}) is non-singular and invertible in the  
range of variation of the field $\f$. However, one can consider  a
function $P$ subjected to stronger (or weaker) conditions, restricting
(or broadening) in this way the notion of conformal equivalence between
2D dilaton gravity models. For example, one can consider as conformally
equivalent models connected by a transformation (\ref{e2}) such that only
those functions $P$, which preserve the structure of the 
singularities of the model, are  allowed. Conversely, one can also allow 
for functions $P$ leading to transformations that become singular at some 
points of the  dilaton field space. Again, an example which has been 
already studied in the literature, can be used to clarify the situation.
Let us consider the models investigated in Ref. \cite {FR}, they are defined by the 
action
\be\lb{g8}
S={1\over 2 \pi} \dx\left[e^{-{2\over n}\f}\left(R+{4\over n}
(\na\f)^2\right) +4\l^{2} e^{-2\f}\right].     
\ee
One can easily show that the models with different values of the parameter
$n$ are conformally equivalent. In fact, by defining 
\be\lb{g9}
D=e^{-{2\over n}\f},\qquad \hgmn=D^{n} \gmn,
\ee 
the action (\ref{g8}) becomes, modulo a total derivative, 
\be \lb{h1}
S={1\over 2\pi}\dhx (DR[\h g] +4\l^2),
\ee
which is the model studied in Ref. \cite {CM3}.
This fact explains way the black hole solutions of the 
models (\ref{g8}) have the same values of mass, temperature and 
Hawking flux as 
the  solution 
of the CGHS model: the black hole solutions of the model (\ref{g8}) are 
related to those of the CGHS model (the model in (\ref{g8}) with $n=1$) 
by a conformal transformation  of the metric. 
However,  the existence and the position of curvature singularities
depend crucially on the parameter $n$.
In fact, the transformation (\ref{g9}) implies 
the following on-shell relation  
\begin{displaymath}
R[g]=D^{n}R[{\h g}] +2nM\l D^{n-2}.
\end{displaymath}
For $0\le n\le 2$ the transformation (\ref{g9}) does not change the 
position of the singularities. The black hole solutions of the
models (\ref{g8}) with these 
values of 
$n$ have a curvature singularity for $D=0$ (the black hole solution 
of the model (\ref{h1}) 
have no curvature singularities but we treat the region of strong 
coupling $D=0$ as a singularity). Conversely, for $n>2$ the position 
of the curvature singularity is changed by the transformation 
(\ref{g9}) from $D=0$ to 
$D=\infty$. In conclusion, if we use the concept of conformal 
equivalence in the restricted sense that only those transformations are
allowed that preserve the structure of  the space-time singularities, 
then only 
transformations (\ref{g9}) with $0\le n\le 2$ are allowed and we have two 
classes of conformally equivalent models, those with $0\le n \le 2$ and
those with $n>2$.    
\begin {flushleft}
{\titolino Acknowledgments}
\end {flushleft}
I took benefit from conversations with S. Mignemi and G. 
Amelino-Camelia. In particular, I am grateful to the latter for having 
drawn my attention to the papers \cite {AS,ABS}.

\end{document}